\documentclass{iau}
\usepackage{graphicx}
\usepackage{natbib}

\title[Jet-induced star formation by the microquasar GRS 1915+105] %% [give here short title] %%
{Jet-induced star formation by a microquasar}

\author[I.F. Mirabel,  S. Chaty, 
 L.F. Rodr{\'i}guez et.al.] %% [give here the short author list; use "et al." if 3 authors or more] %%
{
I. F.  Mirabel$^{1,2}$
,
S. Chaty$^{1,3}$
,
L.F. Rodr\'iguez$^{4}$
\& 
M. Sauvage$^{1}$
}

\affiliation{
$^1$ Laboratoire AIM, CEA/DSM/Irfu/Service d'Astrophysique, Centre de Saclay, B\^at. 709, FR-91191 Gif-sur-Yvette Cedex, France \\ 
email: {\tt felix.mirabel@cea.fr} \\
[\affilskip]
$^2$ Instituto de Astronom\'ia y F{\'i}sica del Espacio C.C. 67, Suc. 28. 1428, Buenos Aires, Argentina \\
$^3$ Institut Universitaire de France, 103, boulevard Saint-Michel, 75005 Paris, France \\
$^4$ Centro de Radioastronom\'ia y Astrof\'isica, UNAM, Campus Morelia, Morelia, Michoac\'an 58190, M\'exico
}

\pubyear{2014}
\volume{313} 
\pagerange{xx--xx}
\jname{Extragalactic jets from every angle}
\editors{F. Massaro, C.C. Cheung, E. Lopez, A. Siemiginowska, eds.}
\begin{document}

\maketitle

\begin{abstract}
Theoretical and observational work show that jets from AGN can trigger star formation. However, in the Milky Way the first -and so far- only clear case of relativistic jets inducing star formation has been found in the surroundings of the microquasar GRS 1915+105. Here we summarize the multiwavelength observations of two compact star formation IRAS sources axisymmetrically located and aligned with the position angle of the sub-arcsec relativistic jets from the stellar black hole binary GRS 1915+105 \citep{mirabel:1994a}. The observations of these two star forming regions at centimeter \citep{rodriguez:1998}, millimeter and infrared \citep{chaty:2001a} wavelengths had suggested -despite the large  uncertainties in the distances a decade ago- that the jets from GRS 1915+105 are triggering along the radio jet axis the formation of massive stars in a radio lobe of bow shock structure.  Recently, \cite{reid:2014} found that the jet source and the IRAS sources are at the same distance, enhancing the evidence for the physical association between the jets from GRS 1915+105 and star formation in the IRAS sources. We conclude that as jets from AGN, jets from microquasars can trigger the formation of massive stars, but at distances of a few tens of parsecs. Although star formation induced by microquasar jets may not be statistically significant in the Milky Way, jets from stellar black holes may have been important to trigger star formation during the re-ionization epoch of the universe \citep{mirabel:2011}. Because of the relative proximity of GRS 1915+105 and the associated star forming regions, they may serve as a nearby laboratory to gain insight into the physics of jet-trigger star formation elsewhere in the universe.      
\end{abstract}

\firstsection % if your document starts with a section,

\section{Observations}
 
The microquasar GRS 1915+105 frequently produces relativistic jets with large kinetic energy, reaching $10^{43}$\,ergs \citep{rodriguez:1999a}. To understand where in the interstellar medium is this energy being dissipated, \cite{rodriguez:1998} carried out radio studies of the surroundings of GRS 1915+105. They found the intriguing possibility of a physical connection of GRS 1915+105 with two infrared/radio sources that appear symmetrically located with respect to GRS 1915+105 and aligned with the position angle of the relativistic jets. 

\begin{figure}
% \vspace*{-2.0 cm}
\begin{center}
 \includegraphics[width=5.3in]{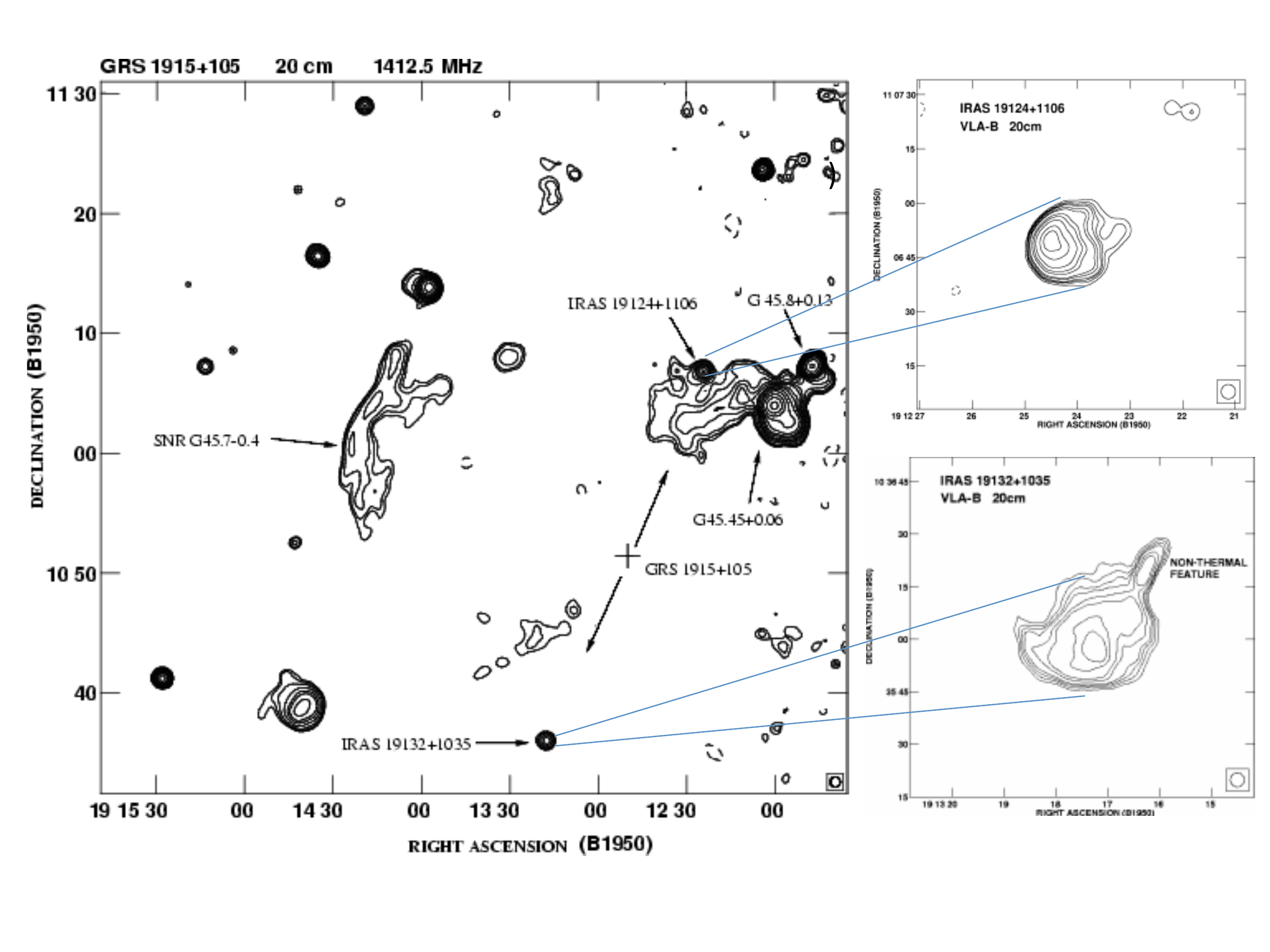} 
% \vspace*{-1.0 cm}
\caption{Left: Map of the surroundings of GRS 1915+105, taken with the VLA-C at 20 cm. The arrows emerging from GRS 1915+105 indicate the position angle of the sub-arcsec relativistic ejecta. Along the axis of the jets from GRS 1915+105 are found two compact infrared/radio sources, IRAS 19124+1106 and IRAS 19132+1035, located at 40-50 parsecs from the microquasar jet source. Contour levels are -3, 3, 4, 6, 10, 15, 20, 30, 40, 60, 100, 200, 400 and 800 mJy/beam. The half power contour of the beam is shown in the bottom right corner. 
Right: VLA maps at 20 cm of the star formation IRAS sources. Note the bow shock structure and non-thermal feature in IRAS 19132+1035 that points to GRS 1915+105. The half power contour of the beam, with diameter of 4'', is shown in the bottom right corner. Contours are -4, 4, 6, 8, 10, 15, 20, 40, 60, 100, 200, 300, and 400 times 0.05 mJy/beam.
}
\label{fig1}
\end{center}
\end{figure}

In Figure \ref{fig1} are shown a mosaic of the surroundings of GRS 1915+105 and the maps  of the two compact infrared/radio sources IRAS 19124+1106 and IRAS 19132+1035 at $\lambda$20 cm obtained with the  VLA. In IRAS 19132+1035 there is to its northwest edge a remarkable feature with a non-thermal spectrum of index -0.8. This feature points back to GRS 1915+105, and as proposed by \cite{rodriguez:1998} and \cite{chaty:2001a}, this non-thermal feature could be produced by the interaction of the jets from GRS 1915+105 and the HII region/compact molecular cloud associated with the IRAS source.   

\begin{figure}
% \vspace*{-2.0 cm}
\begin{center}
 \includegraphics[width=5.3in]{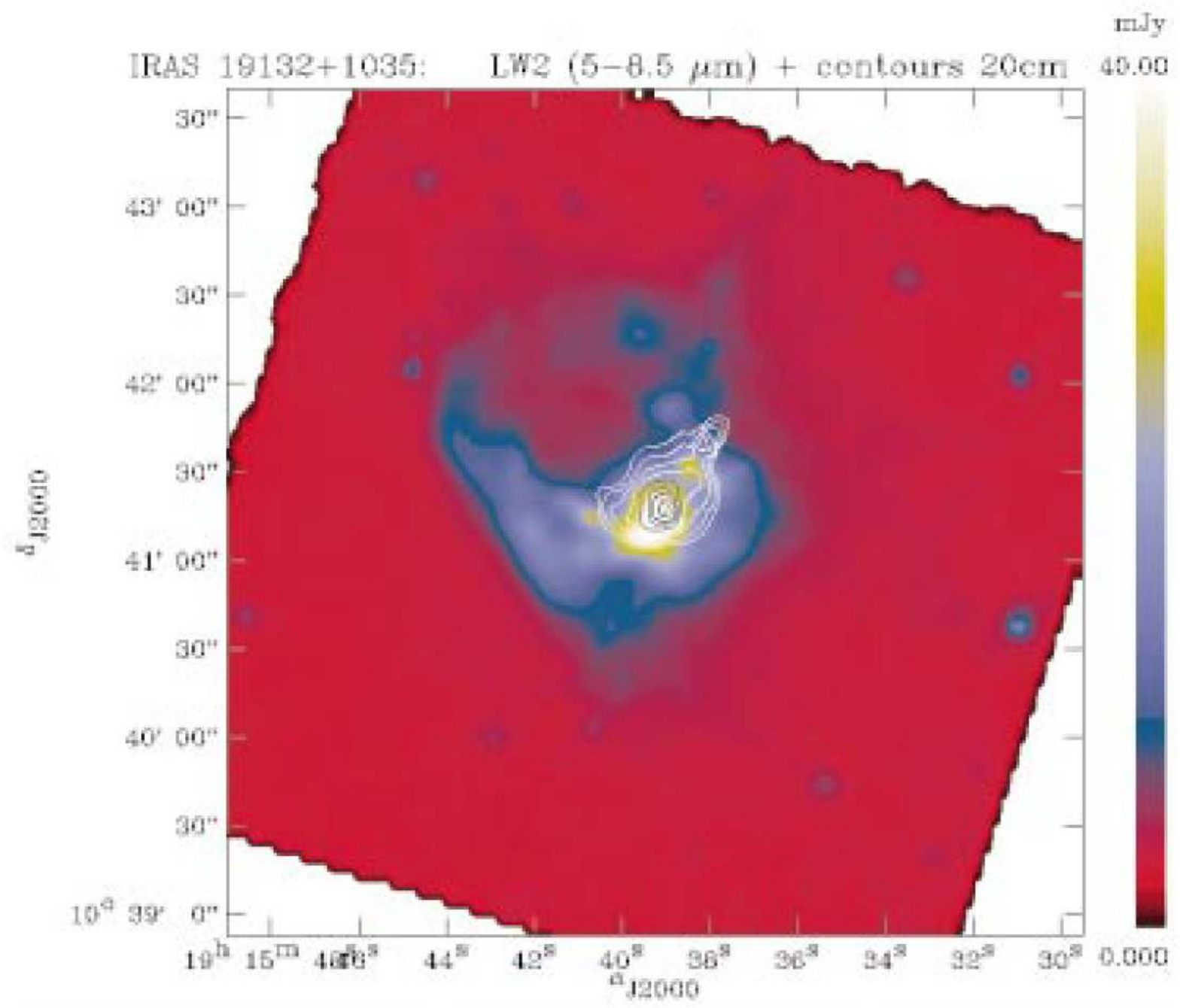} 
% \vspace*{-1.0 cm}
\caption{ISO map of the source IRAS 19132+1035, at $\lambda = 7 \mu m$, taken with the LW2 filter. Superimposed are the 20 cm radio contours at the levels 0.2, 0.4, 0.7, 1.2, 1.8, 2.5, 3, 4 and 5 mJy. Note the bow shock structure in both, the radio and $\lambda = 7 \mu m$ maps. 
}
\label{fig2}
\end{center}

%%%%%%%%%%%%%%%%%%%%%%%%%%%%%%%%%%%%%%%%%%%%%

%%%%%%%%%%%%%%% FIGURE %%%%%%%%%%%%%%%

% \vspace*{-2.0 cm}
\begin{center}
 \includegraphics[width=5.3in]{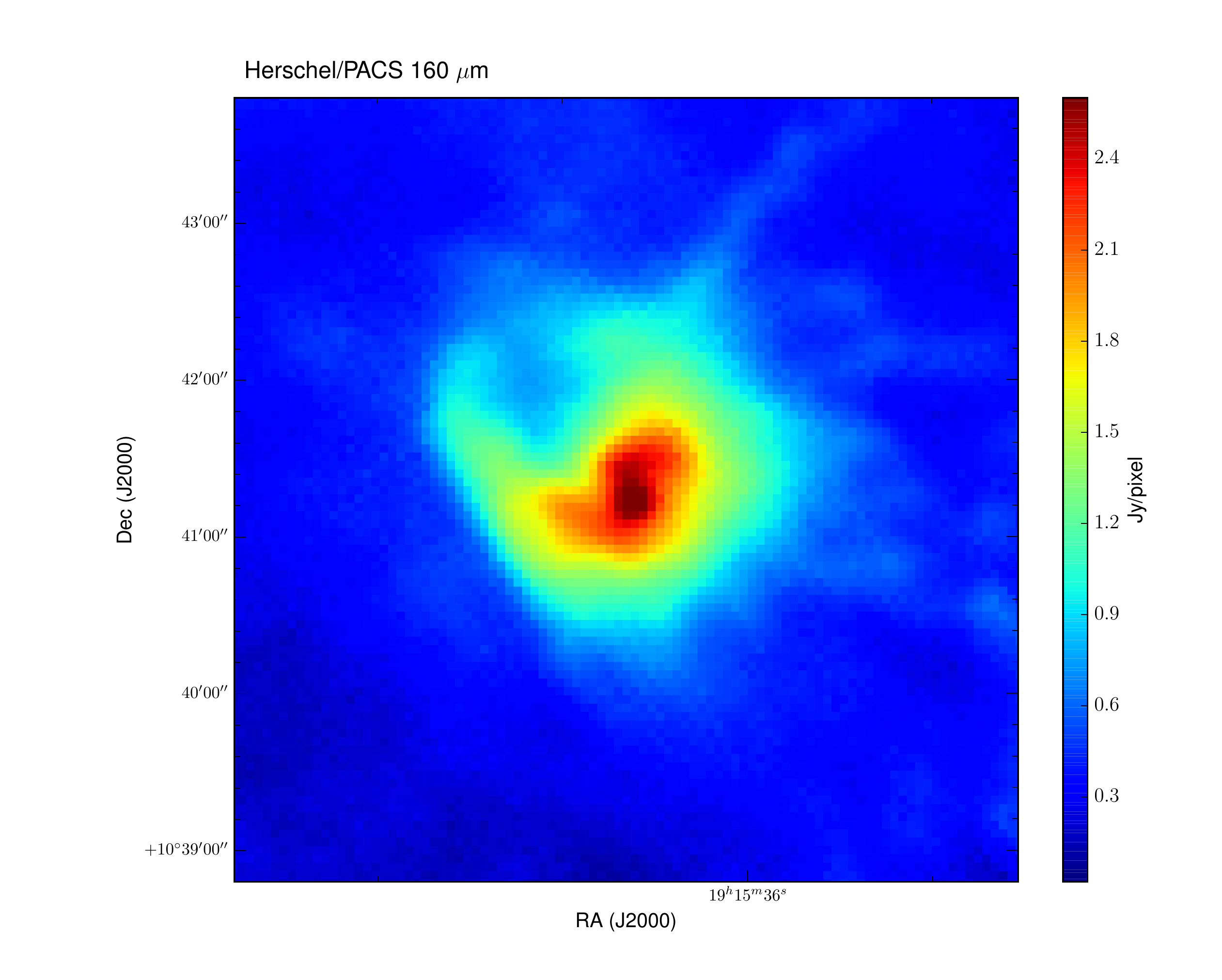} 
% \vspace*{-1.0 cm}
\caption{Herschel  map of IRAS 19132+1035 at $\lambda = 160  \mu m$. The data obtained under the Galactic Plane key project (P.I. S. Molinari) were reduced by one of us. Note the filamentary feature pointing to the NW as the non-thermal feature in the 20 cm maps of Figures 1 and 2. 
}
\label{fig3}
\end{center}
\end{figure}

\cite{chaty:2001a} carried out observations at millimeter (IRAM 30 m), and infrared (ISO, UKIRT, ESO/MPI 2.2 m) wavelengths in both line and continuum emission. Using millimeter tracers of CO, CS and HCO$^{+}$ strong density enhancements are found towards the non-thermal radio feature in the NW of IRAS 19132+1035. At infrared wavelengths strong hydrogen recombination lines and weak lines of molecular hydrogen are observed in that IRAS source. In Figure \ref{fig2} is shown the ISO map at 7 microns, and in Figure \ref{fig3} the Herschel map at 160 microns of IRAS 19132+1035. Both show in the thermal dust emission a similar bow shock structure as seen in the thermal emission at radio waves. 

\section{Evidences for jet induced star formation in GRS 1915+105}

The strongest lines of evidence supporting a physical interaction between the jets from GRS 1915+105 and the compact star forming regions are: (1) the axisymmetric locations of the two compact star forming sources at nearly the same position angle as the sub-arcsec ejections observed at radio wavelengths, (2) the spatial coincidence of the non-thermal radio feature with the inner edge of the southern source IRAS 19132+1035, as well as the orientation of this small non-thermal feature  along the axis of the GRS 1915+105 jets, (3) the location of the highest gas densities in IRAS 19124+1106 and IRAS 19132+1035 on the sides closest to GRS 1915+105 and, in the case of IRAS 19132+1035, close to the non-thermal jet, (4) the bow shock structure in the portion of IRAS 19132+1035 most distant from GRS 1915+105, and (5) the compactness of the molecular clouds associated with the IRAS sources.
   
\section{Discussion}

At the new distance determination of the jet source and IRAS sources of 8.6 (+2.0/-1.6) kpc, and jet inclination with the line of sight of 60° +/- 5° \citep{reid:2014}, the IRAS sources are at ~40-50 pc from GRS 1915+105. Since all these objects are within 1 degree from the Galactic plane, one may ask why there is no triggered star formation closer to GRS 1915+105. Two properties of GRS 1915+105 can account for this: (1) the precession of the jets is less than 10 degrees \citep{rodriguez:1999a}, small enough to remain inside the lobes \citep{chaty:2001a,kaiser:2006}; (2) The accretion to the 12.4 solar mass black hole is by Roche lobe overflow from a red giant star, which may have lasted tens, even hundreds of thousand years, the time for the donor star to cross the Hertzsprung gap and remain in the giant branch. Therefore as proposed by \cite{chaty:2001a}, a large cavity in the form of a tunnel could have been produced, which would explain the ballistic motions of the repeated ejections from GRS 1915+105 observed by \cite{rodriguez:1999a}. 

The high energy of the ejecta, low angular precession of the jet axis, and the possibility that GRS 1915+105 has been an active high energy jet source for tens of thousands years, suggest that the jets from GRS 1915+105 could have overpressured the interstellar gas collecting it and enhancing its density to form a molecular cloud that would ultimatelly collapse forming massive stars. The compactness and displacement of the high density molecular gas associated to the IRAS sources as observed by \cite{chaty:2001a} using several molecular transitions, supports the idea that the jets not only may induce by overpressure the formation of stars, but even induce the formation of the progenitor molecular cloud.  If this is the case, the lobes of GRS 1915+105 and the stars formed in them should be moving away from GRS 1915+105. Our preliminary follow up observations at radio waves suggest in the ionized gas of IRAS 19132+1035 the possibility at 3-4 sigma of proper motions away from GRS 1915+105.

It has been proposed that jet-triggered positive feedback from supermassive black holes produce global star formation, generating and enhancing ultraluminous starbursts with top heavy IMF \citep{silk:2005}. On the other hand, \cite{mirabel:2011} have proposed that due to cosmic chemical evolution, during the re-ionization epoch of the universe the numbers and energy feedback from stellar black hole high-mass X-ray binaries should be greatly enhanced relative to their numbers and energy feedback in the present epoch. Given the much larger densities of the IGM at redshifts greater than z=10, in addition to jets from putative supermassive black holes, jet induced star formation by stellar black holes should have been important to induce star formation. In this context, it will be interesting to probe whether the star formation in IRAS 19132+1035 is top heavy, as predicted by theoretical models. 

%%%%%%%%%%%%%%% FIGURE %%%%%%%%%%%%%%%
%%%%%%%%%%%%%%%%%%%%%%%%%%%%%%%%%%%%%%%%%%%%%

%%%%%%%%%%%%%%% FIGURE %%%%%%%%%%%%%%%
%%%%%%%%%%%%%%%%%%%%%%%%%%%%%%%%%%%%%%%%%%%%%

%%%%%%%%%%%%%%% BIBLIOGRAPHY %%%%%%%%%%%%%%%
\clearpage
%\bibliographystyle{/Users/chaty/Library/Texmf/Bibtex/aa} 
%\bibliography{/Users/chaty/Library/Texmf/Science/science}
%\bibliography{grs1915lobes2.bib}

\begin{thebibliography}{8}
\expandafter\ifx\csname natexlab\endcsname\relax\def\natexlab#1{#1}\fi

\bibitem[{{Chaty} {et~al.}(2001){Chaty}, {Rodr\'{\i}guez}, {Mirabel},
  {Geballe}, \& {Fuchs}}]{chaty:2001a}
{Chaty}, S., {Rodr\'{\i}guez}, L.~F., {Mirabel}, I.~F., {Geballe}, T., \&
  {Fuchs}, Y. 2001, A\&A, 366, 1041

\bibitem[{{Kaiser}(2006)}]{kaiser:2006}
{Kaiser}, C.~R. 2006, MNRAS, 367, 1083

\bibitem[{{Mirabel} {et~al.}(2011){Mirabel}, {Dijkstra}, {Laurent}, {Loeb}, \&
  {Pritchard}}]{mirabel:2011}
{Mirabel}, I.~F., {Dijkstra}, M., {Laurent}, P., {Loeb}, A., \& {Pritchard},
  J.~R. 2011, A\&A, 528, A149+

\bibitem[{{Mirabel} \& {Rodr\'{\i}guez}(1994)}]{mirabel:1994a}

{Mirabel}, I.~F. \& {Rodr\'{\i}guez}, L.~F. 1994, Nature, 371, 46

\bibitem[{{Reid} {et~al.}(2014){Reid}, {McClintock}, {Steiner}, {Steeghs},
  {Remillard}, {Dhawan}, \& {Narayan}}]{reid:2014}
{Reid}, M.~J., {McClintock}, J.~E., {Steiner}, J.~F., {et~al.} 2014, ArXiv
  e-prints

\bibitem[{{Rodr\'{\i}guez} \& {Mirabel}(1998)}]{rodriguez:1998}
{Rodr\'{\i}guez}, L.~F. \& {Mirabel}, I.F. 1998, A\&A, 340, L47, (RM98)


\bibitem[{{Rodr\'{\i}guez} \& {Mirabel}(1999)}]{rodriguez:1999a}
{Rodr\'{\i}guez}, L.~F. \& {Mirabel}, I.F. 1999, ApJ, 511, 398

\bibitem[{{Silk}(2005)}]{silk:2005}
{Silk}, J. 2005, MNRAS, 364, 1337

\end{thebibliography}
%%%%%%%%%%%%%%%%%%%%%%%%%%%%%%%%%%%%%%%%%%%%%

\end{document}